# Heterogeneous Active Messages (HAM) — Implementing Lightweight Remote Procedure Calls in C++

Matthias Noack
noack@zib.de
Zuse Institute Berlin
Berlin, Germany

## ABSTRACT

We present HAM (Heterogeneous Active Messages), a C++-only active messaging solution for heterogeneous distributed systems. Combined with a communication protocol, HAM can be used as a generic Remote Procedure Call (RPC) mechanism. It has been used in HAM-Offload to implement a low-overhead offloading framework for inter- and intra-node offloading between different architectures including accelerators like the Intel Xeon Phi x100 series and the NEC SX-Aurora TSUBASA Vector Engine.

HAM uses template meta-programming to implicitly generate active message types and their corresponding handler functions. Heterogeneity is enabled by providing an efficient address translation mechanism between the individual handler code addresses of processes running different binaries on different architectures, as well a hooks to inject serialisation and deserialisation code on a per-type basis. Implementing such a solution in modern C++ sheds some light on the shortcomings and grey areas of the C++ standard when it comes to distributed and heterogeneous environments.

## 1 INTRODUCTION

Programming applications for distributed, and possibly heterogeneous, systems requires means to control code execution across the participating processes. In traditional HPC applications where large problems are partitioned and distributed across the nodes of a supercomputer or cluster, a Single Program Multiple Data (SPMD) model like MPI can be sufficient. Such SPMD programs typically execute the same binary in each process while a unique identifier for each process, e.g. the MPI rank, controls which process handles which part of the data. More generic scenarios, i.e. Multiple Program Multiple Data (MPMD) programs, require a different model. One such model are Remote Function Calls, that is the possibility to call a function of another process in another address space of the distributed system. Examples for such MPMD applications range from traditional server-client situations, over offload programs for accelerators, to large-scale HPC codes based on frameworks like HPX [7] or Charm++[1]. One way of implementing remote code execution are active messages. In contrast to the notion of messages transferring mere data, active messages are units of execution processing themselves.

In this work, we describe the design and implementation of Heterogeneous Active Messages (HAM), a way of implementing an active message based RPC mechanism in pure C++ that:

- is lightweight,
- generates all active message types and their handlers implicitly via template meta-programming, and
- solves the problem of efficient code address translation between heterogeneous binaries compiled for different platforms.

We explain in detail the caveats of C++ in this context and shed some light on the limitations of the C++ standard when it comes to distributed and heterogeneous computing.

### 1.1 Remote, Distributed, and Heterogeneous

Before continuing, we would like do clarify the use of some terms throughout this paper. Offloading a function to another processor, is synonymous to a remote procedure call (RPC). Remote in RPC refers to any remote address space, not necessarily involving any network transfers. This leads to a very generic interpretation of the term distributed, which from an implementer's point of view would apply to any application running multiple processes that communicate with each other. Each process has its own virtual address space, thus an application using multiple processes is distributed across multiple address spaces. This makes heterogeneous applications, either using some kind of accelerator or coprocessor inside a compute node, or more obviously cluster and supercomputer systems that feature different kinds of processors across their compute nodes, conceptually a subset of distributed applications. The processes of such an application can run on the same or different processors or coprocessors, with or without their own physical memory. They can communicate via shared memory interprocess communication (IPC), PCIe, or some kind of network fabric bridging compute nodes. From the C++ perspective, the notion of address space, process, and thus distributed and heterogeneous is already out of the scope of the current C++17 and prior standards.

The RPC mechanism presented in this work aims to be the most lightweight pure C++ solution usable within an HPC context, and possible other application domains as well. Parallel and distributed applications in such a context usually consist of an somehow orchestrated set of communicating processes that perform a computation together, e.g. a numerical solver, after which the application ends. We do not aim at reliable, long-running services that use RPC-based protocols over the internet and require aspects like versioning and security.

## 2 HAM AND HAM-OFFLOAD

HAM was originally designed as a generic building block for HAM-Offload, a lightweight, C++-only offloading framework. It aims at minimising offloading cost, while still providing a high level of abstraction with support for arbitrary offloading patterns including reverse offloading and offload over fabric. HAM-Offload was introduced and evaluated on the Intel Xeon Phi accelerator with microbenchmarks and a real world application in [16]. It has been recently extended with support for the NEC SX-Aurora TSUBASA [15]. Fig. 2 shows a HAM-Offload example.

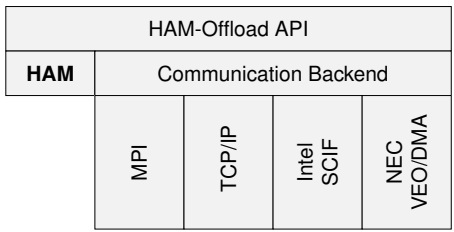

Figure 1: HAM in the context of the HAM-Offload framework. [16]. The Heterogeneous Active Message (*HAM*) mechanism is used to implement the *HAM-Offload API* to offload function calls to other process running on local or remote resource like CPUs or accelerators. The communication between processes is provided by an abstract *Communication Backend*, for which multiple implementations exist.

Fig. 1 shows HAM, which is the scope of this work, in the context of the HAM-Offload architecture. HAM-Offload combines the generic active message capabilities of HAM with an offloading API as a frontend, and different implementations of an abstract communication interface as a backend.

In general, HAM and HAM-Offload can be used on any architecture for which a C++ compiler exists, and between all architectures where one of the supported communication backends is available, and the ABI-constraint, described in Section 5.2, is met. HAM and HAM-Offload were successfully tested between different Intel Xeon and Xeon Phi generations, between Intel Xeon and the NEC SX-Aurora Vector Engine, as well as Intel Xeon and a Cavium ThunderX (ARM64) [4, 15, 16]. The supported communication backends include MPI, TCP/IP, Intel SCIF, and NEC VEO/DMA.

### 2.1 Performance

HAM itself is just a building block that at least requires additional components for communicating the active messages, and a minimal runtime to process and execute them. While the scope of this work is the implementation of an RPC mechanism like HAM in C++ for heterogeneous and distributed systems, HAM has been evaluated as part of HAM-Offload a series of studies.

```cpp
/* HAM-Offload Example */
// inner product of vector a and b
double inner_prod(buffer_ptr<double> a,
                  buffer_ptr<double> b, size_t n) {
    double r = 0.0;
    for (size_t i = 0; i < n; ++i)
        r += a[i] * b[i];
    return r;
}

int main() {
    // host memory
    constexpr size_t n = 1024;
    std::array<double, n> a, b;

    // initialise host memory
    // ...

    // target memory
    node_t target = 1;
    auto a_target = offload::allocate<double>(target, n);
    auto b_target = offload::allocate<double>(target, n);

    // transfer memory
    offload::put(a.data(), a_target, n);
    offload::put(b.data(), b_target, n);

    // async offload, returns a future<double>
    auto result = offload::async(target,
        // function and arguments
        f2f(&inner_product, a_target, b_target, n)
    );

    // do something in parallel on the host
    // ...

    // sync on result future
    double c = result.get();

    return 0;
}
```

Figure 2: A simple HAM-Offload example program that computes the inner product of two vectors. HAM-Offload API elements are highlighted.

Fig. 3 quantifies and compares the offload overhead using HAM-Offload and the corresponding vendor-provided solutions for offloading to an Intel Xeon Phi coprocessor [16] and the NEC Vector Engine coprocessor of a NEC SX-Aurora TSUBASA A300-8 system [15]. On the Intel Xeon Phi, HAM-Offload reduced the overhead of offloading a function to a local accelerator by 28.6×, on the NEC Vector Engine by 13.1×. On the Xeon Phi, this translated into a speedup of up to 2.6× for a real world application. The offload over fabric feature allowed to scale the application transparently use 15 remote accelerators instead of only one or two local ones. The reduced overhead is achieved by the efficient address translation and code execution through HAM, the minimal runtime complexity of HAM-Offload, and on the NEC Vector Engine also by using faster communication paths than the vendor-provided VEO solution [15].

## 3 RELATED WORK

Independent work by other researchers found HAM-Offload to be the best-performing offloading solution in comparison with OpenMP [18], Intel LEO [12], and hStreams [13] for the Intel Xeon Phi coprocessor [6]. An application of HAM-Offload can be found in [9] where the authors implement a domain decomposition solver based on local Schur complements. It uses HAM-Offload to offload a number of dense matrices kernel operations to an Intel Xeon Phi accelerator. Both host CPU and coprocessors are used during individual iterations using a simple load-balancing scheme. Chen et al. [2] included HAM-Offload in a review of programming methods for the Intel MIC coprocessor putting it in the class of high level abstractions.

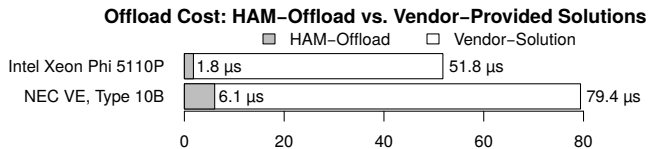

**Figure 3: Comparison of the offload overhead of vendor provided solutions and HAM-Offload measured as the time for offloading an empty function. For the Intel Xeon Phi, Intel LEO [12] is used as vendor solution following the same microbenchmark as published in [16]. For offloading to the NEC Vector Engine (VE), we show the numbers published in [15], measured on an NEC SX-Aurora TSUBASA A300-8 system with NEC VEO as the vendor-provided offloading framework.**

A similar functor-based active message approach without support for heterogeneity can be found in the TACO framework [17], in whose continued development the author was involved. TACO provides the abstraction of a Global Object Space (GOS). HPX is another framework providing such an abstraction, here called an Active Global Address Space (AGAS). Both concepts can be seen as the object oriented flavour of the general Partitioned Global Address Space (PGAS) [3] abstraction. HPX requires all remotely executable functions (called actions) to be explicitly declared as such using macros. While HPX, TACO, and HAM all implement some kind of remote procedure call within the C++ language, other solutions from the HPC context, like Charm++ [1], extend the language and require a special compiler.

## 4 HAM DESIGN
### 4.1 User Perspective

From a user's, i.e. application developer's, perspective we would like to execute a function in the address space of a remote process.

```
int fun(int a, int b) {
    return a + b;
}
```

Ideally, we would like to offload this function as easily as it is to run it asynchronously in a thread:

```
int main() {
    int a, b; // init somehow
    // run asynchronously
    auto res_future =
        std::async(fun, a, b);
    int c = res_future.get();
}
```

Of course, we need to additionally specify to which target process to offload to, and the function call and its arguments need to be transferred into another address space. With HAM, as it is used in HAM-Offload, such an RPC to an offload target would look like this:

```
int main() {
    int a, b; // init somehow
    node_t target; // target process
    // offload asynchronously
    // f2f() generates a closure
    auto res_future =
        offload::async(target,
                       f2f(&fun, a, b);
    int c = res_future.get();
}
```

The offloaded functions or lambdas should declare or use variables with static or thread-local storage only with great care, as they exist per process. Pointers are in general only valid within their original process's address space. HAM-Offload provides special smart pointer types for data handling [15, 16]. Fig. 2 provides a complete example application. In the following we explain the design and implementation of this RPC mechanism, with a focus on the HAM part.

### 4.2 Active Messages and Heterogeneity

As opposed to passive data messages, active messages are executed. The simplest way of implementing an active message would be transferring a function address, that is valid for the receiving process and can be called on reception. In a homogeneous context, this can be achieved by using the exact same executable for all processes that make up the application at runtime. In a scenario of heterogeneous binaries, addresses are only locally valid.

This notion of heterogeneity, based on the binaries used for the communicating processes of the application, starts as soon as different compilers or compilation settings are used. In this sense, a supercomputer that was installed in two phases and features different CPUs, e.g. Intel Xeon processors with the Broadwell microarchitecture featuring AVX2 in phase one, and newer Skylake ones featuring AVX-512 in phase two, can already be seen as a heterogeneous system as the binaries use different instruction sets (although the newer architecture is backwards compatible). A more typical example are systems with heterogeneity within a compute node, like the recent NEC SX-Aurora TSUBASA with an Intel Xeon host featuring multiple NEC Vector Engine PCIe cards, or systems with Xeon Phi accelerators. There are also heterogeneous clusters and supercomputers with different kinds of nodes, like Xeon and Xeon Phi (KNL) nodes, for instance the Cori supercomputer at NERSC, or the Marconi system at CINECA, placed 12 and 19 respectively in the Top-500-list [19]. We exclude GPUs from this study, as they can only execute parts of a program and need a corresponding host program, which would be the target for executing an active message, possibly containing code that uses a local GPU.

The key to implementing heterogeneous active messages is an efficient mechanism to translate the code addresses between processes. The underlying assumption here, is that the code is already there in all processes. This requirement is satisfied by compiling the application for all processes from the same source code, which in an HPC application typically is the case. For HAM, it would be sufficient to make sure the code that entails the template-based code-generation of message types and handler functions is compiled for every generated binary executed by the different processes of the running application.

### 4.3 Structure of HAM

Fig. 4 contains a class diagram showing the main entities of HAM and putting them into relation with each other. Fig. 5 shows a dynamic model of an RPC using the entities at runtime.

The `active_msg_base` class is required as a basic common base class of all active messages. Its only data member is the globally valid handler key. This allows to safely cast every message received as a typeless buffer to this type and call it as a functor with a pointer to that buffer as an argument. The call-operator will then look up the locally valid handler address inside the message handler registry and call it, passing on the pointer to the buffer.

The handler of every message type is a static member function of the `execution_policy` base class template. It knows the actual type of the message which was passed up the inheritance tree via the `Derived` template parameter. This way, the handler can perform an up-cast from the typeless network buffer, back into the type-safe world of C++. The handler is designed as a policy such that a runtime component using HAM can replace it as needed. In its most basic implementation the policy will simply execute the message by calling its call operator, while a more sophisticated runtime might for instance use a policy that puts the message into a queue for a pool of worker threads.

The `active_msg` class template has a static data member that holds the global key for the message handler of its `Derived` type. This serves two purposes. First, it makes sure that, given the type, there is a way to determine the corresponding handler key in $O(1)$, and second, it provides a point at runtime during static initialisation to execute the code that registers the handler address at the message handler registry before the program's `main()`.

Every functor that is safe to be bit-wise copied can inherit the active_msg template and is ready to be used for remote code execution. In Fig. 4, the `offload_msg` template depicts the generic RPC mechanism used in HAM-Offload, which combines HAM's `function` and `active_msg` templates. The `function` template basically is a closure that is constructed from a function, i.e. a signature and a code address, and a set of arguments. The arguments are wrapped in a `migratable` template that provides a hook for serialisation and deserialisation for types that cannot be directly copied. For such a type T, a specialisation of the `migratable` template can specify a converting constructor accepting T for serialisation, and an implicit conversion operator for T performing deserialisation.

At runtime, an exemplary event-chain for an RPC in the HAM-Offload framework looks as shown in Fig. 5. For a given function and its corresponding arguments, a HAM `function` object is created. This functor is then used in HAM-Offload's `async()` function to offload the function call as an `offload_msg`. The receiving process B has to start with a typeless receive buffer, that is first interpreted as an `active_msg_base` instance, whose call operator calls the correct handler using its key data member. The handler performs the upcast into the actual `offload_msg<...>` type which can be executed to generate a result.

The data structures for performing efficient handler address translation at runtime are shown in Fig. 6. Each active message type is associated with a key, allowing the lookup from type to key on the sending side in $O(1)$, while the key acts as an index into the message handler table which allows the lookup of the local handler address on the receiving side in $O(1)$ as well.

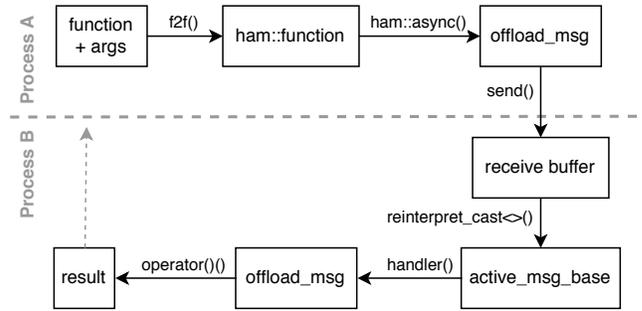

Figure 5: Sequence of entities and transformations for offloading a function call from one process A to process B.

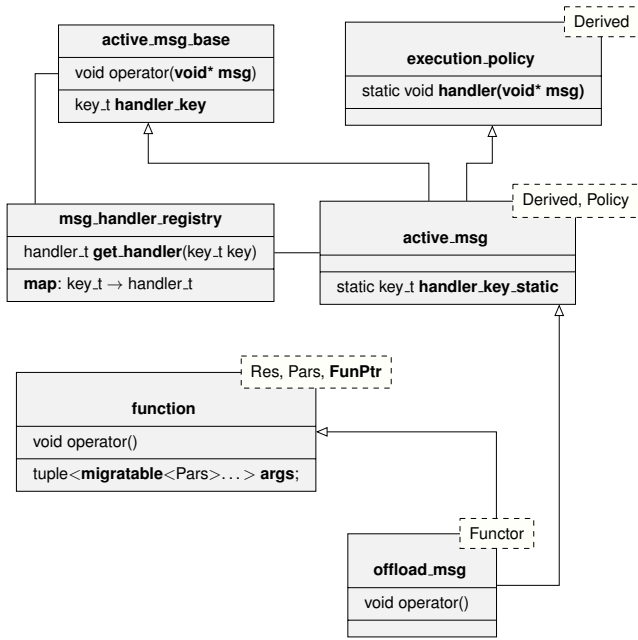

Figure 4: The class hierarchy of HAM. The `offload_msg` combines the handler generation and address translation of the `active_msg` template with a transferable function closure template where complex argument types can be serialised and deserialised by providing a specialisation of the `migratable` template.

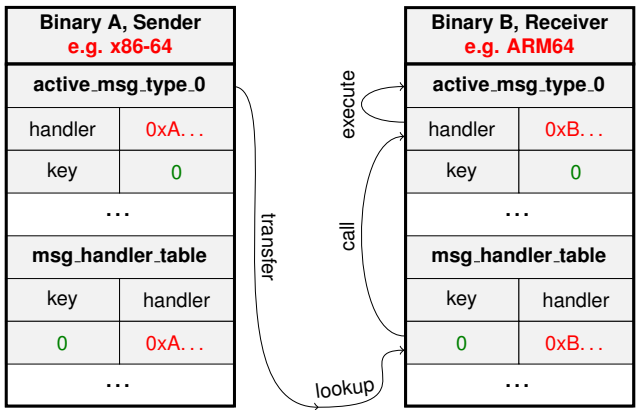

Figure 6: The message-handler address translation between processes. Each active message type has a handler function with a locally valid code address and is associated with the corresponding global key. When transferred to another process, this key is used to look up and call the handler address which performs an upcast from the type-less network buffer to the actual message type and executes it.

## 5 IMPLEMENTATION DETAILS

### 5.1 Functions, Functors, and Lambdas

The function template (see Fig.4) is a generic solution to create transferable closure objects from a function (pointer) and a set of arguments. The outline looks as follows:

```
// function signature as template type parameter
// function pointer as template value parameter
template<typename Result, typename... Pars,
    Result (*FunctionPtr)(Pars...)>
class function<Result (*)(Pars...),
    FunctionPtr> {
public:
    typedef Result result_type;

    // variadic constructor template
    // takes compatible argument types
    template<typename... Args>
    function(Args&&... arguments);

    Result operator()() const;

private:
    // argument storage, uses the actual
    // parameter types, conversion happens
    // on construction
    std::tuple<migratable<Pars>...> args;
};
```

The important part here is, that the function pointer becomes part of the resulting function instantiation's type, not a data member. A function pointer data member would be of no use in the address space of another process. Using the template value parameter makes sure that we get one active message type per function, not per signature. The same function closure type in different processes' binaries is instantiated for different function addresses during compilation. The address translation works on the handler code addresses for each message type. By calling the correct handler, which contains the code to upcast the typeless network buffer back to the actual message type, as explained in Section 4.3, we implicitly get an address translation for the template value argument as well—simply by using the same type in its different code versions across processes and their binaries.

Directly using the function template is somewhat cumbersome, as it would look as follows:

```
function<decltype(fun_ptr), fun_ptr>(
    /* arguments */);
```

Before C++17, a variadic macro construct called f2f() was needed to achieve the following syntax:

```
// f2f = "function to functor"
// NOTE: the '&' is required
f2f(&fun, /* arguments */);
```

With auto template parameters, the following alias can be used instead:

```
template<auto fun_ptr>
using f2f = function<decltype(fun_ptr),
    fun_ptr>;
// C++17 f2f syntax:
// NOTE: the '&' before fun can be skipped
f2f<fun>(/* arguments */);
```

Somewhat challenging pieces of code to offload are those generated by lambda expressions. Capturing lambdas are off limits for distributed systems, as there are no hooks to access or modify the inner state of the generated closure objects. Captureless lambdas however, provide a conversion operator to a function pointer of the corresponding signature type. Since C++17, this operator is constexpr and thus can be used to instantiate templates. The latter is important as the function pointer value is used to initialise a template value parameter (see above). There are limitations to the achievable syntax, though. Lambda expressions themselves cannot be directly used as template *value* arguments, which also prevents using, e.g. std::enable_if and type-traits, to differentiate whether a lambda object or a function pointer was passed to f2f().

```
// NOT possible, lambda used as template argument
f2f<[](/* Pars */){/* do something */}>
    (/* args */);

// possible: unary + operator triggers
// conversion to function pointer
f2f<+[](/* Pars */){/* do something */}>
    (/* args */);
```

The unary + operator can be hidden as follows:

```cpp
// lambda to function
template<typename L, typename Args...>
auto l2f(L lambda, Args&&... args) {
    // conversion to pointer through +
    return f2f<+lambda>(
        std::forward<Args>(args)...);
}
// resulting syntax:
l2f([](/* Pars */){ /* do sth. */ },
    /* args */);
```

While not allowing capturing lambda expressions, this allows to conveniently use anonymous lambdas in-place. For a user of HAM-Offload the code could look in like this example:

```cpp
// offload function
int square(int x) {
    return x * x;
}
// offload functor, f2f as macro (pre C++17)
offload::async(target, f2f(&square, 42));
// offload functor, f2f auto template (C++17)
offload::async(target, f2f<square>(42));

// offload anonymous lambda
offload::async(target, l2f(
    [](int x) { return x * x; },
    42));
```

Another problems with lambda expressions is discussed in the next section.

C++17's user-defined deduction guides for `function` cannot be used to eliminate the `f2f()` and `l2f()` construct, as the function pointer is required as a template value parameter for the instantiation of the `function` template. The pointer value cannot be deduced because the function template generated from a deduction guide's parameter-declaration clause would have to pass it through as a function argument which cannot be `constexpr`, and thus cannot be used as template argument. This could be implemented, if C++ would support `constexpr` in function parameter declarations. There are ways to emulate `constexpr` function parameters by using, e.g. `std::integral_constant` or `constexpr` lambdas, but those would not lead to an improved syntax over the current `f2f()`. Independent of distributed and heterogeneous programming, `constexpr` function parameter declarations in conjunction with template value parameters seem like an interesting feature in general.

As explained in Section 4.3, the `migratable` wrapper used inside the `std::tuple` storing arguments to the offloaded function provides the hooks for serialisation and deserialisation of argument types. HAM provides an additional `is_bitwise_copyable` type-trait template, that can be specialised to flag types as safe or unsafe for unserialised transport between processes. This allows to trigger compiler messages if an unsafe type is used in an offload context.

### 5.2 Message Handler Registry

The `msg_handler_registry` encapsulates the data structures used to collect the handler addresses during static initialisation time before the `main()` function is executed. It uses the *construct on first use* idiom to guarantee initialisation order. The initialisation of the static `active_msg` template member `handler_key_static` performs the registration step for every message type. Each message type registers the handler address, together with the `typeid` name of C++'s RTTI system, and a setter function that is needed later to assign the actual key for the handler. This key is not known at this point of program initialisation, as it is only determined after all handlers have been collected. Hence, the `handler_key_static` is first initialised with a dummy value, that is replaced later by the `msg_handler_registry` calling back the registered setter function.

The second step of the handler registry initialisation requires an explicit call of an `init()` function, typically from the `main()`. Like the registration, this happens inside all processes of the distributed application before any active messages are exchanged. At this point all handler addresses are collected and keys can be assigned. The registry entries were made in an undetermined order during static initialisation. To get them into a deterministic order, they are sorted by the handler names in lexicographical order. The index of each handler in that ordered list becomes the handler key. This way, all processes end up with the same key-to-handler mapping without any communication during initialisation.

However, this builds on the underlying assumption that `typeid` returns the same names in all processes, or at least names that imply the same lexicographical order across processes. The `typeid` data is implementation specific, i.e. not defined in the C++ standard. Most compilers return the mangled name of the type, which is defined in the ABI, which is not standardised.

In practise, this turns out not to be an actual problem, as most compilers and systems use ABIs [8, 10] that reference the Intel IA-64 (Itanium) ABI [5] for the type name mangling convention. The ABI used by the NEC compiler for the Vector Engines of the NEC SX-Aurora TSUBASA does not specify an ABI yet [11], but apparently uses the same convention. So far, we have never witnessed an incompatibility throughout our tests. In the actual HAM code base, `msg_handler_registry` is an interface implemented by `msg_handler_registry_abi`. This way, it could be easily exchanged—if required—by a different registration mechanism, not using RTTI, but for instance user-specified identifiers.

Fig. 7 shows an exemplary output of the handler map (`typeid` name to address) and indexed handler vector (global handler key to address). Even though the `function` template does contain a function pointer as template value parameter, the name of the corresponding function is internally used. This works in our favour as the names stay equal in different binaries, but reveals a problem with lambdas: the compiler's internally used names for lambda functions are reflected in the `typeid` name and they do not adhere to any standard or convention. Different compilers use different names: GCC for instance uses `_FUN`, while LLVM/Clang names the corresponding function `__invoke`. Hence, programs using active messages derived from lambda expressions work only if the same compiler is used for all binaries, or two compilers happen to use the same name.

```
=================== BEGIN HANDLER MAP =====================
typeid_name:
  N3ham3msg10active_msgINS_7offload6detail11offload_msgINS2_7runtime17
      terminate_functorENS0_23execution_policy_directEEES7_EE
handler_address: 0x440d10
typeid_name:
  N3ham3msg10active_msgINS_7offload6detail18offload_result_msgINS_8functionIPFiiEXadL_ZZ13ham_user_mainiPPcEN3$_08
      __invokeEiEEEENS0_24default_execution_policyEEESC_EE
handler_address: 0x42a7e0
typeid_name:
  N3ham3msg10active_msgINS_7offload6detail18offload_result_msgINS_8functionIPFvvEXadL_Z7
      fun_onevEEEENS0_24default_execution_policyEEES9_EE
handler_address: 0x42db20
=================== END HANDLER MAP =======================
=================== BEGIN HANDLER VECTOR ==================
index: 0,   handler_address: 0x440d10
index: 1,   handler_address: 0x42a7e0
index: 2,   handler_address: 0x42db20
=================== END HANDLER VECTOR ====================
```

**Figure 7: An output of the actual handler tables generated by HAM in the context of HAM-Offload. The first handler is an internally used terminate message used to communicate the end of the program, the second entry is the result of an offloaded lambda expression, and the last one represents a simple void fun_one() function offload.**

## 6 LIMITATIONS OF THE C++ STANDARD

Even though C++ has no language support (yet) for distributed or heterogeneous systems, or applications consisting of more than one process or more than one address space, a lot of functionality can be implemented as libraries where the low-level details are conveniently hidden behind high-level abstractions. Free functions and member functions can be handled through address translation mechanisms. Data types of different kinds can be handled using wrappers, and serialisation if necessary. Partitioned global address or object spaces can be created across processes by using smart-pointers that combine an address space or process identifier with a local pointer. However, the implementation of HAM has shown some caveats and limitations of C++ which might be resolvable in future versions of the standard.

Less of a practical problem, but definitely a problem in formal arguments is the fact that the `type_info` returned by `typeid` holds implementation-specific information. In HAM, the name member, that usually holds the mangled name of the queried type, is used to generate the message handler table assuming that the `typeid` names of all processes will share the same lexicographical order. For all tested compilers and architectures, that assumption was never proven wrong, but it remains something that needs to be tested or derived from implementation specifications like the ABI in addition to the C++ standard. While a completely standardised C++ ABI might put too many constraints on implementers which might affect performance on different architectures, a standardised `type_info` in case RTTI (RunTime Type Information) is supported by a compiler could be argued for.

Another language feature that is troublesome in distributed and heterogeneous contexts are lambda expressions. There are no hooks to access the functor objects generated from lambda expressions in a way that would allow transferring them between address spaces. Only the trivial case of a captureless lambda expressions can be handled as shown in Section 5.1, but constraints compatibility between binaries created by different compilers. Harmonising the observable names for lambda functions inside compiler-generated code seems like a rather low hanging fruit. For capturing lambdas, even if serialisation and deserialisation functionality for the captured types were implemented, there are no hooks to use them. At least for lambdas capturing by copy, this could be interesting for practical implementations of distributed applications. Lambda expressions capturing by reference, lead to another, very general problem of many non-distributed languages like C++, the fact that there are no language means to qualify parameters as input, output, or both. Only the `const` qualifier can be used to rule out output-semantics for a function argument. This missing functionality can be partly solved by generic wrappers that can be evaluated to avoid unnecessary data processing and transfers in RPCs.

A similarly generic problem are implementation-dependent types. This affects basic types like `int` and alike, whose size differs on architectures of different word size. The best solution is to avoid them, and use types of defines size like `int64_t` instead. A similar case is the `long double` type which can be sometimes found in numerical HPC applications. Depending on the architecture, it can be anything from a synonym for double over an extended precision format, up to a quadruple precision number. If it must be communicated such a type requires serialisation and deserialisation.

While there are standard proposals to enhance the reflection capabilities of C++ beyond `typeid`, it is important that the implementations of the resulting language features will be compatible across heterogeneous binaries as well. Basically everything that is required to implement distributed and heterogeneous applications, but is implementation defined, i.e. outside the C++ standard is a potential problem for interoperability across different compilers and architectures used to build and run a distributed or heterogeneous application. For future C++ standards, an effort in standardising these interoperability-relevant parts of the language without constraining compiler implementers in performance critical areas would be a major step towards programming distributed and heterogeneous systems.

## 7 CONCLUSION

We presented HAM, a C++ library for lightweight Remote Procedure Calls (RPC) in heterogeneous and distributed applications. The design and implementation were discussed in detail, showing how to implement an efficient message handler address translation mechanism, as well as a generic functor template for closure objects that can be safely communicated between processes. HAM is used in the HAM-Offload framework, that has been shown to reduce the offloading overhead by factors of 28.6× and 13.1× on the Intel Xeon Phi coprocessor and NEC Vector Engine, respectively, as compared to the vendor provided solutions.

The implementation of HAM explores the boundaries of the C++ standard with respect to the possibilities of enabling heterogeneous and distributed computing. While many modern C++ features are crucial or at least helpful in implementing a lightweight RPC mechanism such as HAM, there are also features that are lacking, while others are deemed to stay in the the homogeneous world of their local address space. One such example are lambda expressions that expose compiler-specific naming conventions through their RTTI information, and do not offer any hooks to access the internal state of the resulting objects for serialisation and deserialisation. For non-lambda types, name-mangling conventions used in the RTTI are practically standardised through the ABI specifications, most of which referencing the IA-64 ABI [5].

For future standardisation efforts, reviewing the implementation-defined parts of the C++ standard with respect to their effect on interoperability between compilers and architectures, as well as a potential performance-loss by constraining compiler implementers seems like a logical next step towards distributed and heterogeneous computing with C++. HAM and HAM-Offload are free open source software and available on GitHub [14].

## ACKNOWLEDGMENTS

We thank Thomas Steinke, Marius Knaust, and the anonymous reviewers for their valuable comments that helped to improve this paper, as well as Marcel Erhardt for the fruitful discussions and feedback on HAM, especially regarding C++17. The author acknowledges the North-German Supercomputing Alliance (HLRN) for providing access to the supercomputing resources operated at ZIB. This work is partially carried out within the IPCC activities at ZIB with support from Intel, and partly supported by the Deutsche Forschungsgemeinschaft (DFG) in the framework of the Priority Programme "Software for Exascale Computing" (SPP-EXA), DFG-SPP 1648, project FFMK (Fast Fault-tolerant Microkernel).

## REFERENCES

[1] Bilge Acun, Abhishek Gupta, Nikhil Jain, Akhil Langer, Harshitha Menon, Eric Mikida, Xiang Ni, Michael Robson, Yanhua Sun, Ehsan Totoni, Lukasz Wesolowski, and Laxmikant Kale. 2014. Parallel Programming with Migratable Objects: Charm++ in Practice. In *Proceedings of the International Conference for High Performance Computing, Networking, Storage and Analysis (SC '14)*. IEEE Press, Piscataway, NJ, USA, 647–658. https://doi.org/10.1109/SC.2014.58

[2] C. Chen, F. Yang, F. Wang, L. Deng, and D. Zhao. 2018. Review of Programming and Performance Optimization on CPU-MIC Heterogeneous System. In *2018 IEEE 3rd International Conference on Image, Vision and Computing (ICIVC)*. 894–900. https://doi.org/10.1109/ICIVC.2018.8492841

[3] Mattias De Wael, Stefan Marr, Bruno De Fraine, Tom Van Cutsem, and Wolfgang De Meuter. 2015. Partitioned Global Address Space Languages. *ACM Comput. Surv.* 47, 4, Article 62 (May 2015), 27 pages. https://doi.org/10.1145/2716320

[4] Daniel Deppisch. 2018. *Advancing the Heterogeneous Active Messages Approach*. Master's thesis. Humboldt-Universität zu Berlin, Faculty of Mathematics and Natural Siences, Department of Computer Science.

[5] Intel Corporation [n. d.]. *Itanium C++ ABI, v1.86*. Intel Corporation.

[6] Jeffers, James and Reinders, James. 2013. *Intel Xeon Phi Coprocessor High Performance Programming* (1st ed.). Morgan Kaufmann Publishers Inc., San Francisco, CA, USA.

[7] H. Kaiser, M. Brodowicz, and T. Sterling. 2009. ParalleX An Advanced Parallel Execution Model for Scaling-Impaired Applications. In *Parallel Processing Workshops, 2009. ICPPW '09. International Conference on*. 394–401. https://doi.org/10.1109/ICPPW.2009.14

[8] H.J. Lu, Milind Girkar, Michael Matz, Jan Hubika, Andreas Jaeger, and Mark Mitchell. [n. d.]. *System V Application Binary Interface, K1OM Architecture Processor Supplement, v1.0*.

[9] Malý, Lukáš and Zapletal, Jan and Merta, Michal and Čermák, Martin. 2018. Xeon Phi acceleration of domain decomposition iterations via heterogeneous active messages. *AIP Conference Proceedings* 1978, 1 (2018), 360004. https://doi.org/10.1063/1.5043963 arXiv:https://aip.scitation.org/doi/pdf/10.1063/1.5043963

[10] Michael Matz, Jan Hubika, Andreas Jaeger, and Mark Mitchell. [n. d.]. *System V Application Binary Interface, AMD64 Architecture Processor Supplement, Draft v0.99.6*.

[11] NEC Corporation [n. d.]. *System V Application Binary Interface VE Architecture Processor Supplement, v1.1*. NEC Corporation. https://www.nec.com/en/global/prod/hpc/aurora/document/VE-ABI_v1.1.pdf

[12] ChrisJ. Newburn, Rajiv Deodhar, Serguei Dmitriev, Ravi Murty, Ravi Narayanaswamy, John Wiegert, Francisco Chinchilla, and Russell McGuire. 2013. Offload Compiler Runtime for the Intel Xeon Phi Coprocessor. In *Supercomputing*. Springer Berlin Heidelberg, 239–254. https://doi.org/10.1007/978-3-642-38750-0_18

[13] C. J. Newburn, G. Bansal, M. Wood, L. Crivelli, J. Planas, A. Duran, P. Souza, L. Borges, P. Luszczek, S. Tomov, J. Dongarra, H. Anzt, M. Gates, A. Haidar, Y. Jia, K. Kabir, I. Yamazaki, and J. Labarta. 2016. Heterogeneous Streaming. In *2016 IEEE International Parallel and Distributed Processing Symposium Workshops (IPDPSW)*. 611–620. https://doi.org/10.1109/IPDPSW.2016.217

[14] Matthias Noack. 2019. HAM-Offload GitHub Repository. (2019). https://github.com/noma/ham

[15] Matthias Noack, Erich Focht, and Thomas Steinke. 2019. Heterogeneous Active Messages for Offloading on the NEC SX-Aurora TSUBASA. In *2019 IEEE International Parallel and Distributed Processing Symposium Workshops (IPDPSW)*.

[16] Matthias Noack, Florian Wende, Thomas Steinke, and Frank Cordes. 2014. A Unified Programming Model for Intra- and Inter-node Offloading on Xeon Phi Clusters. In *Proceedings of the International Conference for High Performance Computing, Networking, Storage and Analysis (SC '14)*. IEEE Press, Piscataway, NJ, USA, 203–214. https://doi.org/10.1109/SC.2014.22

[17] Jörg Nolte, Yutaka Ishikawa, and Mitsuhisa Sato. 2001. TACO: Prototyping High-Level Object-Oriented Programming Constructs by Means of Template Based Programming Techniques. *SIGPLAN Not.* 36 (December 2001), 35–49. Issue 12. https://doi.org/10.1145/583960.583965

[18] OpenMP Architecture Review Board 2018. *OpenMP Application Program Interface, Version 5.0*. OpenMP Architecture Review Board. https://www.openmp.org/wp-content/uploads/OpenMP-API-Specification-5.0.pdf

[19] TOP500.org. 2018. Top500: TOP 500 Supercomputer Sites. (November 2018). http://www.top500.org